\documentclass[letterpaper]{article} 
\usepackage{aaai2026}  
\usepackage{times}  
\usepackage{helvet}  
\usepackage{courier}  
\usepackage[hyphens]{url}  
\usepackage{graphicx} 
\usepackage{natbib}  
\usepackage{caption} 
\usepackage{algorithm}
\usepackage{algorithmic}

\usepackage{amsmath}
\usepackage{amssymb}

\usepackage{newfloat}
\usepackage{listings}
\DeclareCaptionStyle{ruled}{labelfont=normalfont,labelsep=colon,strut=off} 
\floatstyle{ruled}
\newfloat{listing}{tb}{lst}{}
\floatname{listing}{Listing}
\title{Developmental Symmetry-Loss: A Free-Energy Perspective \\on Brain-Inspired Invariance Learning}
\author {
    Arif D\"onmez
}
\affiliations{
    Leibniz Research Institute for Environmental Medicine (IUF)\\
    arif.doenmez@ruhr-uni-bochum.de
}

\begin{document}

\maketitle

\begin{abstract}
    We propose \textbf{Symmetry-Loss}, a brain-inspired algorithmic principle that enforces invariance and equivariance through a differentiable constraint derived from environmental symmetries. 
    The framework models learning as the iterative refinement of an \emph{effective symmetry group} $\hat{G}$, paralleling developmental processes in which cortical representations align with the world’s structure. 
    By minimizing structural surprise, i.e. deviations from symmetry consistency, Symmetry-Loss operationalizes a \emph{Free-Energy--like objective} for representation learning. 
    This formulation bridges predictive-coding and group-theoretic perspectives, showing how efficient, stable, and compositional representations can emerge from symmetry-based self-organization. 
    The result is a general computational mechanism linking developmental learning in the brain with principled representation learning in artificial systems.
\end{abstract}

\section{Motivation}

Brains do not learn by passively fitting data; they develop internal models that reflect the symmetries and regularities of the world. 
From early visual processing to higher cognition, neuronal populations gradually acquire invariances---such as translation, reflection, and compositional structure---that stabilize perception and support generalization \citep{DiCarlo2012,Bertamini2018,Fumarola2022}. 
This developmental process can be interpreted as a progressive \emph{alignment of internal symmetries} with environmental dynamics \citep{Higgins2022,Cooray2024}.

Recent theoretical frameworks, including the \emph{Free-Energy Principle} and \emph{predictive coding}, describe this alignment as the minimization of surprise or prediction error \citep{Friston2010,Parr2019}. 
Similarly, symmetry-based representation learning in machine intelligence \citep{Higgins2022} and cognitive-developmental theories \citep{Lake2017,Tenenbaum2011} both emphasize structure discovery through transformation consistency.

In this work, we propose \textbf{Developmental Symmetry-Loss}, a differentiable mechanism that operationalizes this principle within modern learning systems. 
The method penalizes deviations from group-consistent transformations and iteratively refines an \emph{effective symmetry group} $\hat{G}$, which stabilizes representational dynamics. 
Conceptually, this procedure mirrors how cortical representations specialize through symmetry preservation and symmetry breaking during development \citep{Fumarola2022,Keck2025}.

By interpreting symmetry consistency as a form of \emph{structural free-energy minimization}, our approach links predictive-coding theories in neuroscience with invariant learning in machine intelligence \citep{Friston2010,Whittington2019}. 
The resulting framework provides both (i) a theoretical bridge between biological and artificial learning and (ii) a practical mechanism for stable, energy-efficient, and interpretable representation formation. 

In summary, \emph{Developmental Symmetry-Loss} proposes that learning invariances through iterative symmetry refinement is not merely an engineering strategy---but an algorithmic reflection of how the brain itself learns to model its world.

\section{Symmetry Refinement: Theory and Formulation}

Learning in biological and artificial systems can be viewed as the process of aligning internal transformations with the symmetries of the environment. 
We formalize this idea through a differentiable objective that penalizes representational deviations from group-consistent transformations. 

\subsection*{Prerequisites}
\paragraph{Foundations: Actions, invariants.}
Let $G$ denote a candidate topological group, and $\mathcal{F}$ a topological $G$-modul. We denote by 
$\mathcal{F}^G := \{  f \in \mathcal{F} \,:\, g.f=f ~~~\forall g\in G \}$ the zeroth cohomology group of $G$ with coefficients in $\mathcal{F}$. We call its elements \textit{symmetries}. In the present work, we are only interested in the case where $\mathcal{F}$ is some class 
of functionals of a \textit{latent} space $\mathcal{X}$, i.e. $\mathcal{F} = \mathcal{F}(\mathcal{X})$. 

We use the basic algebraic geometry setting \citep{shafarevich1994basic}. Although algebraic geometry is not essential for understanding the framework, it provides a solid and rigorous foundation that ensures the formulation remains technically accurate.
Let the latent space $\mathcal{X}\subseteq\mathbf{A}^n:=\mathbf{K}^n$ be an affine variety over $\mathbf{K}\in\{ \mathbf{R}, \mathbf{C} \} $, and let $G$ be an algebraic group that acts polynomial on $\mathcal{X}$. 
This means $\mathcal{X}$ is a $G$-variety. In this context, we are interested in the class of \textit{polynomial functionals} also known as regular functionals. We denote these functionals as $\mathcal{F} = \mathbf{K}[\mathcal{X}] = \{ f:\mathcal{X}\to \mathbf{K} \,:\, f \; \mathrm{is\;regular} \}$.

The action of $G$ on $\mathcal{X}$ makes $\mathcal{F}$ to a topological $G$-module, defined by $(g \cdot f)(x) := f(g^{-1}x)$, where $g \in G$, $f \in \mathcal{F}$, and $x\in \mathcal{X}$ \citep{kraft1985geometrische}. 
We are particularly interested in finding symmetries in $\mathcal{F}$, i.e. $\mathcal{F}^G \subseteq \mathcal{F}$. 
In other words, we seek polynomial functionals of the latent space that remain invariant under the action of $G$. The ring structure of $\mathcal{F}^G$ is inherited by $\mathcal{F}$, also called the \textit{invariant ring}.
\paragraph{Geometric invariant theory quotient.}\label{catquot}

From this point forward, let $G$ be a linear reductive group. When $G$ acts on a latent space $\mathcal{X}$, the first step is to examine the quotient map $\pi : \mathcal{X} \to \mathcal{X} // G$ induced by the action. 
Understanding this map is essential, as it reveals the structural properties of the action itself.

In the polynomial setting, the quotient map takes the form of the \textit{geometric invariant theory quotient} (GIT quotient):
$$
\pi: \mathcal{X} \to \mathcal{X} // G,\;x \mapsto \left(\chi_1(x),\ldots, \chi_m(x)\right),
$$
where $\mathcal{X} // G := \mathrm{Spec}\left(\mathcal{F}^G\right) \subseteq \mathbf{A}^m$.

This is a \textit{categorical quotient} and the functionals $\chi_1, \ldots, \chi_m \in \mathcal{F}^G$ form a \textit{complete fundamental system of symmetries}, meaning that $\mathbf{K}[\chi_i : i = 1, \ldots, m] = \mathcal{F}^G$. 
Notably, the function $\pi$ serves to identify orbit closures; that is, for any points $v, x \in \mathcal{X}$, we have $\overline{Gx} = \overline{Gv}$ if and only if $\pi(x) = \pi(v)$ \citep{kraft1985geometrische,mumford1994geometric}.

This elegantly illustrates how symmetries of $\mathcal{F}$ are linked to the group's action on the domain space.

\subsection{Target-free symmetry refinement and disentanglement}
\paragraph{Symmetry-Loss functional.}
Let $\Omega$ denote the input/sample space, $\mathcal{X}$ a latent space and $\phi_\theta : \Omega \rightarrow \mathcal{X}$ be a parameterized representation map. 

Let $G$ denote a candidate group of transformations acting on the latent space $\mathcal{X}$, and let $\chi_1, \ldots, \chi_m\in\mathcal{F}(\mathcal{X})$ be its complete fundamental system of symmetries.

For $v_0\in\Omega$ let 
\begin{equation*}
    \left(\chi_1(\phi_\theta(v_0)), \ldots, \chi_m(\phi_\theta(v_0)) \right) =: \left(y_1,\ldots, y_m\right)\in\mathbf{A}^m.
\end{equation*}
Since the $\chi_i$'s identify orbits (closures), deviations from this relation quantify a violation of \emph{orbit-alignment}.

We therefore define the \emph{Symmetry-Loss} relative $v_0$ as
\begin{equation*}
    \mathcal{L}_{\mathrm{sym}}(\phi_\theta; G,v_0)
    =  \sum_{i=1}^m\,\mathbf{E}_{\omega\sim \mathcal{D}}\Big[\Vert\, y_i - \chi_i\left(\phi_\theta(\omega)\right) \Vert^2_2\Bigg],
    \label{eq:sym_loss}
\end{equation*}
where $\mathcal{D}$ is the empirical data distribution over $\Omega$, and $G$ the current hypothesis of environmental symmetries. If the context is clear, we will drop the $v_0$ in the notation.

Minimizing $\mathcal{L}_{\mathrm{sym}}$ (with respect to the model parameters $\theta$) enforces local consistency relative $v_0$ of representations under the group action, analogous to minimizing prediction error in predictive-coding theories. In settings lacking a fixed anchor element $v_0$, the symmetry--loss is minimized jointly over the model parameters and the latent variables $y_i$, allowing the system to self--organize reference points and thereby eliminate dependence on an externally chosen anchor.

We further define the \textit{orbit-loss} of $x\in\mathcal{X}$ relative $v_0$ as  
\begin{equation*}
    \mathcal{L}_{\mathcal{O}, v_{0}}(\phi_\theta; G,x)
    =  \sum_{i=1}^m\,\Vert\, y_i - \chi_i\left(x\right) \Vert^2_2.
    \label{eq:sym_loss}
\end{equation*}

\paragraph{Structural surprise.}
We interpret $\mathcal{L}_{\mathrm{sym}}$ as a measure of \emph{structural surprise}:
the mismatch between the symmetry predicted by the model relative $v_0$ and the symmetry expressed by the data distribution $\mathcal{D}$.
Whereas classical free-energy minimization reduces sensory surprise, symmetry-loss reduces structural surprise by aligning internal transformations with external regularities.

\vspace{0.8em}
\noindent
The preceding discussion used algebraic–geometric language (varieties, invariant rings, GIT quotients) to clarify how group actions induce symmetry coordinates.
In what follows we work in the smooth category: all spaces are smooth manifolds and maps are diffeomorphisms (or locally diffeomorphic in practice).
Conceptually, the two regimes are parallel:
\begin{itemize}
    \item[(i)] invariant generators $(\chi_i)$ provide orbit–separating coordinates in the algebraic case, while in the smooth case one may use separating invariants, sufficient statistics, or learned surrogates in the symmetry–loss;
    \item[(ii)] categorical quotients $\mathcal{X}\!//G$ correspond to orbit spaces modulo $G$ in the smooth setting (up to regularity);
    \item[(iii)] group actions, intertwiners, and orbit–alignment have identical formal roles.
\end{itemize}
The GIT intuition (quotients via invariants) is retained as a guiding picture, but all optimization and refinement steps are carried out in the smooth category.

\paragraph{Iterative Disentanglement by Symmetry Refinement.}
In the absence of explicit targets, learning reduces to organizing
representations so that they respect the intrinsic symmetries of the data.
Rather than aligning outputs to fixed labels, the model aligns internal
transformations to the structural regularities of the input space itself.
This defines a \emph{target–free} or \emph{self–supervised} setting, in which
the learning signal arises purely from deviations from orbit consistency.

Consider a representational space $\Omega \subset \mathbf{A}^n$ whose samples are subject to latent transformations. 
We assume that the observed variability in $\Omega$ arises from the (possibly partial or unknown) action of a group $G$. 
Our aim is to iteratively uncover a coordinate system in which this action becomes increasingly disentangled. 

Let $\Gamma_0 = \Omega \times \Omega$ denote the initial relation between inputs and their current representations. 
Given an input group $G_0$ acting on $\Omega_1\times\Omega$, we learn a diffeomorphism 
$\phi_0 : \Omega \to \Omega_1$ such that the transformed graph 
\[
    \Gamma_0' = \{\, (\phi_0(x),\, x) \mid x \in \Omega \,\} 
       \subseteq  \Omega_1\times\Omega
\]
approximates a single orbit of $G_0$, i.e. 
\[
    \Gamma_0' \approx G_0 \cdot (\phi_0(x_0),\, x_0),
\]
by minimizing $\mathcal{L}_{\mathrm{sym}}(\phi_0; G_0)$ for orbit consistency. 
After minimizing, a residual misalignment typically remains, indicating unmodelled latent symmetries. 
We therefore introduce a new latent space $\Omega_2$, a group $ G_1 $ acting on $\Omega_2\times\Omega$, 
and a further coordinate refinement $\phi_1 : \Omega_1 \to \Omega_2$, 
again minimizing $\mathcal{L}_{\mathrm{sym}}(\phi_1; G_1)$, such that the transformed graph
\[
    \Gamma_1' = \{\, (\phi_1(\phi_0(x)),\, x) \mid x \in \Omega \,\}
       \subseteq  \Omega_2\times\Omega
\]
approaches a single orbit of $G_1$. 
Repeating this yields a hierarchy
\[
    (\phi_k, G_k)_{k \geq 0}, \qquad \Omega_{k+1} = \phi_k(\Omega_k),
\]
which defines an \emph{iterative disentanglement process}. 
Each step refines the representation by reparametrizing coordinates 
so that the new latent manifold becomes increasingly symmetric.

\vspace{0.3em}
\noindent
\paragraph{Intertwiners and effective symmetries.}
Each refinement step defines an equivariant morphism
\[
(\phi_k\times\mathrm{id}):(\Omega_k\times\Omega,\,G_k^{(k)})
   \longrightarrow (\Omega_{k+1}\times\Omega,\,G_k),
\]
where $G_k^{(k)}$ denotes the \emph{induced action} of $G_k$ on
$\Omega_k\times\Omega$ obtained by conjugation through
$\phi_k\times\mathrm{id}$:
\[
G_k^{(k)}
   := (\phi_k\times\mathrm{id})^{-1}\,G_k\,(\phi_k\times\mathrm{id})
   \subset \mathrm{Diff}(\Omega_k\times\Omega).
\]
This definition ensures that each map $\phi_k\times\mathrm{id}$ is an
\emph{intertwiner} between the group actions
$(G_k^{(k)},\,\Omega_k\times\Omega)$ and $(G_k,\,\Omega_{k+1}\times\Omega)$,
satisfying
\[
(\phi_k\times\mathrm{id})\circ g^{(k)} = g\circ(\phi_k\times\mathrm{id}),
\qquad g\in G_k.
\]

For any earlier index $j<k$, the corresponding action of $G_j$ can be
transported to the level $\Omega_k\times\Omega$ via repeated conjugation,
\[
G_j^{(k)}
   := (\phi_{k-1}\times\mathrm{id})\cdots
      (\phi_{j+1}\times\mathrm{id})\,
      G_j\,
      (\phi_{j+1}\times\mathrm{id})^{-1}\cdots
      (\phi_{k-1}\times\mathrm{id})^{-1},
\]
so that all induced actions are expressed in the same coordinate frame.
The cumulative (or \emph{effective}) symmetry at step~$k$ is then defined
as the closure of all these transported actions,
\[
\widehat G_k :=
   \big\langle\, G_0^{(k)},\,G_1^{(k)},\,\dots,\,G_{k-1}^{(k)} \,\big\rangle
   \subseteq \mathrm{Diff}(\Omega_k\times\Omega).
\]
Hence, each refinement augments the effective group by adjoining the new
induced action $G_k^{(k)}$, yielding the extension
$\widehat G_{k+1}=\langle\,\widehat G_k,\,G_k^{(k+1)}\,\rangle$.
This process captures how orthogonal, previously unmodelled symmetries
become internalized during successive refinements.

\vspace{0.5em}
\noindent
The effective groups form an ascending chain of inclusions
\[
\widehat{G}_0
   \subseteq \widehat{G}_1
   \subseteq \widehat{G}_2
   \subseteq \dots,
\quad
\widehat{G}_\infty = \bigcup_{k\ge0}\widehat{G}_k
   \subset \mathrm{Diff}(\Omega_\infty\times\Omega).
\]
Equivalently, the system
$\{(\Omega_k\times\Omega,\,\widehat G_k)\}_{k\ge0}$
forms a direct (inductive) system of $G$–spaces.
Its colimit
\[
\widehat{G}_\infty
   = \varinjlim_k \widehat{G}_k
\]
represents the \emph{emergent symmetry} acting on the limiting relation manifold.
The sequence of intertwiners
\[
(\phi_k\times\mathrm{id})
   : (\Omega_k\times\Omega,\,\widehat G_k)
   \longrightarrow (\Omega_{k+1}\times\Omega,\,\widehat G_{k+1})
\]
thus defines a coherent diagram of equivariant representations.
The limiting group $\widehat G_\infty$ captures the
\emph{symmetry closure induced by the hypothesized groups}
$\{G_k\}$—that is, the largest symmetry consistently representable
within the chosen refinement sequence.
When the input groups coincide with the true latent symmetries,
$\widehat G_\infty$ recovers the full orbit structure of the data;
otherwise it yields an approximate, model-dependent effective symmetry
reflecting the inductive biases imposed during training.

In this way, the iterative refinement may be viewed as an algebraic
growth process of symmetry representations, whose accuracy and
expressiveness depend on the adequacy of the hypothesized groups.

The colimit view highlights how refinement accumulates symmetry structure.
To make this process explicit, we next examine the algebraic relation
between consecutive effective groups $\widehat G_k$ and $\widehat G_{k+1}$.
Each step can be interpreted as an \emph{extension} that measures how
previously unmodelled, approximately orthogonal symmetries become integrated
into the effective representation.

\paragraph{Group extensions and orthogonality.}
At each iteration, the refinement augments the effective symmetry
by adjoining the newly induced action $G_k^{(k)}$,
yielding
\[
\widehat G_{k+1}
   = \langle\, \widehat G_k,\, G_k^{(k)} \,\rangle
   \subseteq \mathrm{Diff}(\Omega_{k+1}\times\Omega).
\]
This refinement induces an extension of effective symmetry actions.
When $\widehat G_k$ is normal in $\widehat G_{k+1}$, this extension can be
represented by a short exact sequence
\[
1 \rightarrow \widehat G_k
\longrightarrow \widehat G_{k+1}
\longrightarrow
\Delta G_k
\rightarrow 1,
\qquad
\Delta G_k := \widehat G_{k+1}/\widehat G_k,
\]
where $\Delta G_k$ measures the additional symmetry degrees of freedom
introduced at refinement step $k$.

In the general case, $\Delta G_k$ should be interpreted as an
\emph{effective quotient} capturing new orbit directions modulo the
previously internalized symmetries; formally, it may be viewed as a
quotient in the category of group actions (equivalently, as a
homogeneous space of $\widehat G_{k+1}$ modulo $\widehat G_k$),
rather than as an abstract group.

When $\Delta G_k$ acts independently of $\widehat G_k$
(i.e.~their intersection is trivial),
the refinement is said to be \emph{orthogonal},
and $\widehat G_{k+1}$ forms a semi–direct product
$\widehat G_{k+1} \simeq \widehat G_k \rtimes \Delta G_k$.
Otherwise, the groups overlap nontrivially, indicating that the newly
added symmetry partially reuses or reparametrizes degrees of freedom
already represented at earlier layers.

Hence, the refinement hierarchy can be understood as a sequence of
nested symmetry extensions,
\[
1 \to \widehat G_0
   \to \widehat G_1
   \to \widehat G_2
   \to \cdots,
\]
whose structure reflects how new symmetries are discovered and
internalized during learning.
Orthogonal extensions correspond to genuinely novel latent factors,
whereas overlapping ones signal redundancy or coupling among the
symmetry components already acquired.

In practice, orthogonal extensions correspond to the emergence of
independent latent factors, each contributing a distinct and separable
mode of variability within the representation.
Non–orthogonal extensions, in contrast, indicate partial entanglement,
where the newly introduced symmetry overlaps with or modulates those
already internalized.
From this perspective, iterative symmetry refinement can be viewed as a
process of progressive factorization of the representation manifold,
in which successive layers isolate increasingly independent symmetry
components of the data–generating process.
Whether this progressive separation succeeds depends on the adequacy of
the hypothesized symmetry sequence~$G_k$ and on the learnability of the
associated intertwiners.
When these conditions align, the refinement gradually decomposes composite
transformations into independent symmetry factors, realizing
disentanglement as an emergent structural property rather than a guarantee.

\subsection{Regression via orbit completion}
The previous subsection described the target-free case, where refinement
is driven solely by hypothesized group symmetries.
We now show how the same mechanism applies when an explicit mapping
$f:\Omega\!\to\!Y$ is provided, yielding a geometric formulation of regression.
When a target function $f:\Omega\to Y$ is given, its graph itself defines
the relation manifold on which symmetry refinement operates.


\paragraph{Orbit completion and prediction.}
Fix a target function $f:\Omega\to Y$ and its graph
\[
\Gamma_f=\{(x,f(x))\}\subset \Omega\times Y.
\]
Under refinement, we transport the graph by
\[
\begin{array}{rcl}
    \Gamma_k' \; & = & \;\{\,(\Phi_k(x),\,f(x))\mid x\in\Omega\,\}\;\subset\;\Omega_{k+1}\times Y, \\[.5em]
    \Phi_k & = & \phi_k\circ\cdots\circ\phi_0,
\end{array}
\]
and hypothesize a group $G_k$ acting on $\Omega_{k+1}\times Y$ such that
$\Gamma_k'$ is (approximately) a single $G_k$–orbit.
Let $a_k:=(\Phi_k(x_\ast),\,f(x_\ast))$ be an anchored training pair and write
\[
\mathcal{O}_k \;:=\; G_k\cdot a_k \;\subset\; \Omega_{k+1}\times Y.
\]
The refinement loss enforces $\Gamma_k'\approx \mathcal{O}_k$
(\emph{orbit alignment}).
Given a novel input $x_0$, prediction amounts to completing the pair
$(\Phi_k(x_0),\,y)$ so that it lies (as well as possible) on $\mathcal{O}_k$:
\[
   \hat y_k(x_0)\;\in\;\arg\min_{y\in Y}\;\mathcal{L}_{\mathcal{O}, a_{k}}\bigr(\Phi_k; G_k, (\Phi_k(x_0),\,y) \bigr)   
\]
Thus $y$ is the ``missing puzzle piece'' that places $(\Phi_k(x_0),y)$
on the current orbit.

\paragraph{Consistency across refinement.}
As orbit alignment improves with $k$, $\mathcal{O}_k$ becomes an increasingly
faithful proxy for $\Gamma_k'$, so that $\hat y_k(x_0)$ stabilizes.
Under local transversality (the graph being a local section to the
$G_k$–orbits) the completion is unique; otherwise, the minimum-norm or
regularized solution is selected among admissible $y$'s.

Regression is realized \emph{implicitly}:
one learns $\Phi_k$ through symmetry refinement, and for a new input $x_0$,
predicts $y$ by \emph{orbit completion}.
In this view, generalization emerges as the geometric continuity
of orbit alignment across unseen samples.

\paragraph{Remark.}
This perspective places supervised regression and unsupervised
representation learning within the same algebraic geometry:
both seek to embed data into a manifold of consistent orbits.
Whereas the target-free case discovers stable orbits internally,
the supervised case anchors them externally via the graph of~$f$,
revealing prediction as the act of completing a symmetry relation.

\section{Theoretical Examples}

\paragraph{Classes of actions on the latent space.}
The general orbit–alignment principle subsumes classical invariance and
equivariance as special cases of how the group acts on the
target–latent product.
If the group $G$ acts trivially on the target component,
$g\!\cdot\!(x,y)=(gx,y)$, the orbit condition enforces
$\phi(gx)=\phi(x)$ and the symmetry–loss reduces to the familiar
invariant representation objective.
If the target itself transforms under a representation
$\rho:G\!\to\!\mathrm{Aut}(Y)$, i.e.\ $g\!\cdot\!(x,y)=(gx,\rho(g)y)$,
orbit alignment becomes $\phi(gx)=\rho(g)\phi(x)$, recovering the
classical notion of equivariance.
More generally, when the transformation law couples input and output
nontrivially—e.g.\ through a semi–direct or twisted product
$1\!\to\!H\!\to\!G\!\to\!Q\!\to\!1$—the relation takes the form
\[
\phi(q\!\cdot\!x)=\rho(q)\phi(x)\tau(q,x),
\]
where the cocycle $\tau:Q\times\Omega\!\to\!H$ encodes the interaction
between subgroups.
Such ``twisted'' actions arise naturally when multiple latent factors
interact hierarchically or multiplicatively.
The refinement framework accommodates all these cases uniformly:
each refinement step models a new extension
$\widehat G_{k+1}=\langle\widehat G_k,G_k^{(k+1)}\rangle$,
thereby integrating additional coupling terms into the effective
symmetry group without altering the underlying learning principle.

\paragraph{Remark.}
The additional factor $\tau(q,x)\!\in\!H$ acts as a \emph{cocycle correction}
arising from the nontrivial extension
$1\!\to\!H\!\to\!G\!\to\!Q\!\to\!1$.
It encodes how transformations in the quotient $Q$ modulate those in the
subgroup $H$.
When $\tau\!\equiv\!e$, the law reduces to standard equivariance;
a nontrivial $\tau$ corresponds to a \emph{twisted equivariance}
in which input transformations induce context–dependent modulations of
the output.

\section{Implementation and Computational Scalability}

From a computational standpoint, each map $\phi_k$ is instantiated as a learnable neural \emph{intertwiner}—typically a multilayer perceptron—trained to realize the corresponding group action through the proposed symmetry–loss. The key enabler of this framework is the formulation of the loss in terms of a \emph{complete fundamental system of symmetries} $(\chi_i)$, which provides an invariant coordinate system on the latent space. Optimization thus proceeds entirely in these invariant coordinates and does not require explicit evaluation or sampling of group transformations. This property makes the approach both scalable and adaptable: the group structure is captured implicitly through the algebraic dependencies among the $\chi_i$, rather than through explicit enumeration of its elements or representations. Consequently, the framework retains full architectural freedom—any differentiable model, such as a multilayer perceptron, convolutional network, or graph–based encoder, can serve as the intertwiner $\phi_k$. Training uses standard gradient–based optimizers (e.g.\ Adam or SGD) without modification, since the gradients are defined entirely through differentiable invariants. In this way, the refinement hierarchy becomes a \emph{neural ladder of intertwiners}: a compositional stack of learned coordinate transformations that progressively internalize the symmetry structure of the data, combining theoretical expressiveness with computational efficiency.

\section{Related Work and Conceptual Connections}

The proposed framework builds upon and extends several converging lines of research on symmetry, representation learning, and neural computation.

This work can be viewed as a natural continuation of the initial orbit-alignment framework introduced in \citet{donmez2023discovering}, which offered a first proof-of-concept but contained no refinement mechanism.

Early advances in disentangled representation learning emphasized the role of structured latent spaces that capture independent factors of variation \citep{higgins2018towards,achille2018life}. 
These approaches introduced the notion of \emph{latent homologies}—correspondences between domains that share abstract relational structure—which closely parallels the present view of orbit alignment as matching transformations across representational levels.
However, existing methods typically rely on explicit sampling or reconstruction-based objectives, whereas the present formulation leverages \emph{fundamental symmetries} identified through invariant generators, leading to a symmetry–loss that enforces geometric consistency without requiring enumeration over group elements.

More recent theoretical work has explicitly connected symmetry-based representation learning with cognitive and neurobiological theories of general intelligence \citep{Higgins2022}.
In this view, both artificial and biological systems achieve generalization by progressively internalizing the symmetries of their environment—a principle mirrored in the iterative refinement mechanism developed here.
Our framework can thus be interpreted as a differentiable implementation of this symmetry-based learning process.

Within neuroscience, predictive-coding and free-energy principles provide complementary accounts of how the brain minimizes surprise by maintaining symmetry-consistent internal models \citep{friston2009free, Cooray2024, Fumarola2022}.
The refinement steps in our model parallel hierarchical inference in cortical circuits, where successive layers align internal transformations with increasingly abstract invariances.
This correspondence situates the proposed algorithm at the interface between geometric machine learning and computational neuroscience.

Finally, the developmental perspective articulated by Lake and colleagues \citep{Lake2017}—that human learning proceeds through successive refinement of compositional and causal representations—offers a cognitive analogue to our iterative symmetry refinement: each $\phi_k$ learns to ``re-align'' internal coordinates with latent generative factors, akin to the child’s progressive discovery of invariant structure in perception and action.
Together, these connections position the present framework as a unified geometric account of representation learning that spans artificial and biological domains.

\section{Neurobiological Interpretation and Discussion}

The iterative symmetry–refinement framework proposed here can be interpreted
as a computational analogue of representational development in the brain.
Biological learning does not begin with explicit supervision but through
progressive internal reorganization of sensorimotor mappings to reflect the
symmetries and regularities of the environment.
From early visual processing to higher cognition, neural populations gradually
acquire invariances—such as translation, reflection, or compositional structure—
that stabilize perception and enable generalization
\citep{DiCarlo2012,Bertamini2018,Fumarola2022}.
This process can be viewed as a gradual alignment of internal symmetry groups
to the dynamical symmetries of the external world
\citep{Higgins2022,Cooray2024}.

Theoretical frameworks such as the \emph{Free–Energy Principle} and
\emph{Predictive Coding} formalize this alignment as the minimization of
surprise or prediction error \citep{Friston2010,Parr2019}.
Within this view, our proposed refinement algorithm corresponds to a
\emph{structural} form of prediction minimization: rather than adjusting
expectations about sensory states, it adapts the very symmetry relations
that organize those expectations.
Each refinement step thus parallels a neurodevelopmental process in which
the brain learns to reorganize its internal representational geometry
toward more abstract, factorized, and compositional forms.

This interpretation resonates with developmental theories of
\emph{intuitive physics} and \emph{learning–to–learn}
\citep{Lake2017,marcus2020next}.
Infants appear to construct successively refined internal models that
capture stable transformations in their sensorimotor world.
In the same way that children infer latent causal variables underlying
observed dynamics, each refinement step in our framework discovers
latent symmetry generators that explain observed transformations.
Analogously, each refinement step in our framework can be viewed as a
learned “update of inductive biases,” internalizing new symmetry factors
while preserving previously stabilized ones.
In this sense, symmetry refinement provides a computational account of
the gradual emergence of structured, symmetry–based intelligence—
bridging neural development, representation learning, and theoretical
neuroscience.

\section{Outlook: Adaptive Symmetry Discovery}

The present framework establishes symmetry refinement as a differentiable process
that progressively internalizes structural regularities of the data.
A natural next step is to couple this refinement mechanism with reinforcement
learning principles, enabling the system to \emph{actively explore} and
\emph{hypothesize} new symmetry transformations \citep{caselles2019symmetry}.
Rather than fixing the candidate groups $\{G_k\}$ a priori,
an agent could adjust its symmetry hypotheses based on predictive success or
expected free–energy minimization, thereby learning \emph{which} transformations
yield stable orbits.
This integration would extend the current model from passive symmetry alignment
to an \emph{active symmetry discovery} paradigm, in which perception, prediction,
and action jointly shape the refinement trajectory.
In such a view, representation learning, control, and curiosity become unified
under the same organizing principle: the active pursuit of symmetry in the
structure of experience.

From a neurobiological perspective, such a mechanism resonates with
developmental learning in infants.
Early sensorimotor exploration can be interpreted as a process of testing and
refining internal symmetry hypotheses about the world—
for instance, discovering invariances under translation, rotation, or causal
reversal through embodied interaction \citep{Lake2017,friston2019dynamic, Higgins2022}.
Each refinement step corresponds to an incremental reorganization of neural
representations toward more abstract and compositional invariants,
mirroring the progressive emergence of intuitive physics and structured
world models in early cognition.
Extending the proposed refinement algorithm with reinforcement signals
would thus move it closer to the biological paradigm of
\emph{learning to learn}—where the discovery of symmetries is guided not
only by reconstruction accuracy but also by the intrinsic drive to
reduce uncertainty and maintain predictive control over the environment.


\section{Acknowledgments}
The author thanks his wife and children for their support and inspiration.

\bibliography{aaai2026}


\end{document}